\documentstyle[prl,aps,multicol]{revtex}
\input epsf
\begin{document}
\draft
\title{Dynamics and Configurational Entropy in the LW Model for
Supercooled Orthoterphenyl}
\author{S.~Mossa$^{1,2}$, E.~La Nave$^{1}$, H.~E.~Stanley$^{1}$, 
C.~Donati$^2$, F.~Sciortino$^2$ and P.~Tartaglia$^2$}
\address{
$^1$ Center for Polymer Studies and Department of Physics,
Boston University, Boston, Massachusetts 02215\\
$^2$Dipartimento di Fisica, INFM and INFM Center 
for Statistical Mechanics and Complexity,\\
Universit\`a di Roma "La Sapienza" ,
Piazzale Aldo Moro 2, I-00185, Roma, Italy
}
\date{\today}
\maketitle
\begin{abstract}
We study thermodynamic and dynamic properties of a rigid model of the
fragile glass forming liquid orthoterphenyl. 
This model, introduced by Lewis and Wahnstr\"{o}m in 1993,
collapses each phenyl ring to a single interaction 
site; the intermolecular site-site interactions are 
described by the LJ potential whose parameters
have been selected to reproduce some bulk properties
of the orthoterphenyl molecule. A system of $N=343$ molecules
is considered in a wide range of densities and
temperatures, reaching simulation times up to 1 $\mu s$.
Such long trajectories allow us to equilibrate
the system at temperatures below the mode coupling temperature
$T_c$ at which the diffusion constant reaches values 
of order $10^{-10} \:cm^2/s$ and thereby to sample in a significant 
way the potential energy landscape in the entire temperature range.
Working within the inherent structures thermodynamic formalism,
we present results for the temperature and density dependence
of the number, depth and shape of the basins of the potential energy
surface. We evaluate the total entropy of the system
by thermodynamic integration from the ideal --non interacting-- 
gas state and the vibrational entropy approximating the 
basin free energy with the free energy 
of $6N-3$ harmonic oscillators.
We evaluate the configurational part of the entropy as a difference
between these two contributions.
We study the connection between thermodynamical and dynamical
properties of the system. We confirm that the temperature dependence of the 
configurational entropy and of the diffusion constant,
as well as the inverse of the characteristic structural 
relaxation time, are strongly connected in supercooled 
states; we demonstrate that this connection is well 
represented by the Adam-Gibbs relation, stating a linear 
relation between $\log D$ and the quantity $1/T S_c$.
This relation is found to hold both above and below
the critical temperature $T_c$ ---as previously found in the case of 
silica--- supporting the hypothesis that a connection exists
between the number of basins and the connectivity properties
of the potential energy surface.
\end{abstract} 

\pacs{PACS number(s):}
\begin{multicols}{2}

\section{INTRODUCTION}
\label{intro}
Understanding the dynamic and thermodynamic 
properties of supercooled liquids
is one of the more challenging tasks of condensed
matter physics (for recent reviews see 
Refs.~\cite{debenedetti01,mezard01,tarjus00,debenedettibook}
and references therein). A significant amount of 
experimental~\cite{cummins,cummins-pisa,gotze-pisa,angell95,torre}, 
numerical~\cite{kob-review} 
and theoretical work~\cite{mct1,mct2,mezpar,wolynes,speedy} 
is being currently devoted to the understanding of the physics 
of the glass transition and to the associated slowing down 
of the dynamics.
Among the theoretical approaches, an important role has been
played by the mode coupling theory (MCT)~\cite{mct1,mct2}, which,
interpreting the glass transition as a purely dynamical 
phenomenon, has constituted a significant tool for the interpretation
of both experimental\cite{cummins,torre,feyer,expt1,expt2,expt3} 
and numerical simulation results\cite{kobgleim,fabbian,sciortinokob} 
in weakly supercooled states. 

In the last years the study of the topological
structure of the potential energy (hyper-) surface (PES)
~\cite{stillinger_pes} and the connection between the properties 
of the PES and the dynamical behavior 
of glass forming liquids has become an active field of 
research. Building on the inherent structure (IS) 
thermodynamic formalism proposed long
time ago by Stillinger and Weber~\cite{stillinger_pes},  
the PES can be uniquely partitioned in local basins and 
properties of the basins explored in supercooled states 
(average basin depth and basin volume), have been quantified.
Studies have mainly focused on two fundamental questions:
{\em (i)} which are the basins relevant for the thermodynamics
of the system, i.e., which are the basins 
populated with largest probability?
and {\em (ii)} which are the topological properties of the
regions of the PES actually explored by the system
during its dynamics?
From this point of view, the PES approach has somehow
unified, at least on a phenomenological level, the
thermodynamic and dynamic approaches to the glass transition.

Numerical analysis of the PES has shown that
trajectories in configuration space can be separated into intra-basin
and inter-basin components \cite{schroder,harrowell}.
 The time scales of the two components become increasingly 
separated on cooling.
The intra-basin motion has been associated with  the high-frequency
vibrational dynamics, while the structural relaxation
($\alpha$-relaxation) has been related  to the
process of exploration of different basins. 
It has also been shown that on lowering $T$, 
the system  populates  basins  of lower and lower 
energy\cite{sastry00}. 
The $T$ dependence of the depth of the typical sampled 
basins follows a $1/T$ law~\cite{sastry01,starr01,heuer}
for fragile liquids, and, for strong liquids, it appears 
to approach a constant value on cooling~\cite{voivod01}. 
The number of basins $\Omega$ as a function of the basin depth 
$e_{IS}$  has also been recently evaluated for a few models
\cite{lennard_jones_pes,scala00,starr01,voivod01,speedyjpc,sastry01}, 
opening the possibility of calculating the so-called 
configurational entropy $S_c$ and its $T$ dependence. $S_c$, defined as
the logarithm of the number of accessible basins 
$S_c \equiv k_B \log \Omega$,  has been successfully 
compared with theoretical predictions\cite{mezpar,coluzzi}.
At the same time, the approaches and the 
techniques developed for
the analysis of the PES of structural glasses have spread to
the field of disordered spin systems, where similar
calculations have been performed\cite{crisanti01} and
similar conclusions have been reached.
The evaluation of $S_c$ for models of glass-forming liquids
allows to numerically check, in a very consistent 
way,  the relation between $S_c$ and the systems 
characteristic time $\tau$, proposed by 
Adam and Gibbs~\cite{adam65},
and recently ``derived'' in a novel way\cite{wolynes}. 
Numerical support for a relation between the
$T$ dependence of $S_c$ and the $T$ dependence of $\tau$, 
although limited to very few models, 
is providing new physical insight on the connection
between thermodynamics and long time dynamical properties.
The ideas developed within the inherent structure formalism  
have also been generalized to out-of-equilibrium conditions where 
the slow aging dynamics has been interpreted as
the process of searching for basins of increasingly deep energy  
~\cite{aging_pes1,aging_pes2,aging_pes3,stjpcm01}.

In this paper we study the properties of the PES for a rigid
model (LW) of the fragile glass former orthoterphenyl, first 
introduced  by Lewis and Wahnstr\"{o}m~\cite{lewis} 
and recently revisited by Rinaldi {\it et al.}\cite{rinaldi01}.
We have studied the properties of the PES in a 
temperature range in which the diffusion coefficient
varies by more than four orders of magnitudes for five different
density values. This work attempts to build a bridge between 
models of more direct theoretical interest, like 
Lennard Jones (LJ) and soft spheres, and models which appear
to reproduce, even if in a crude way, properties of 
complex materials. In this respect, orthoterphenyl is the best
candidate, being one of the most studied glass 
forming liquids~\cite{expt1}.
The LW model is a three-sites model, with intermolecular
site-site interactions described by the 
LJ potential. 
This model is among the simplest models for a nonlinear
molecule. The limitation constituted by the fact that 
it does not take into account the internal molecular degrees of
freedom (see~\cite{otp_flex} for a more realistic model), 
is overruled by the observation that its simplicity
---it can be considered as an atomic LJ with constraints---
allows one to reach simulation times of the order of  
$\mu s$. Hence a significant sampling of the PES
in a large temperature and density range is possible. 
Moreover, this model constitutes an ideal bridge between 
simple atomic models and molecular models, being possible to 
treat it under several approximations\cite{rinaldi01}. 

The paper is structured as follows:
In Sec.~\ref{configurational_entropy}
we briefly recall the main results of the IS formalism. 
In Sec.~\ref{configurational_entropy_2} we show 
the calculation of the configurational entropy
as a difference between the total entropy and the vibrational entropy.
In Sec.~\ref{numerics} we give some numerical details.
We present our results in Sec.~\ref{results}, which is divided 
into subsections detailing the calculation 
of the total entropy by thermodynamical integration
from the ideal gas state, the study of the 
vibrational properties of the PES and the calculation 
of the configurational entropy. In the end we study the link 
between configurational entropy and the diffusion constant,
investigating the validity of the Adam-Gibbs equation.
In Sec.~\ref{conclusions} we finally discuss our results 
and we draw some conclusions.
In Appendix A we report the analytical 
calculation of the total entropy of a system of LW molecules 
in the non-interacting ``ideal gas'' limit. 

\section{Inherent Structure Thermodynamics formalism}
\label{configurational_entropy}
In this section we briefly review the IS formalism in the
NVT ensemble~\cite{stillinger_pes,jpcm00}, 
the extension to the NPT ensemble poses no particular 
problems~\cite{stillinger_pes}. This formalism has become
an important tool in the numerical analysis of
classical models since it is numerically possible to calculate
in a very precise  way the inherent structures 
(defined as the local minima of the PES) 
and hence compare the theoretical predictions with the
numerical results. Given an instantaneous configuration of the system,
a steepest descent path along the potential energy hypersurface 
defines the  closest IS.

In the IS formalism, the partition function of a system is written as 
a sum over all the PES basins. Basins of given IS energy
contribute non-negligibly to the total sum  
if their IS energy is very low, if their volume is very large 
and/or if they are highly degenerate, i.e. 
several basins are characterized by this IS energy. 
This corresponds to partition the phase space  
in the local energy minima of the PES 
and their basins of attraction. Such a partition 
is motivated by the fact that in supercooled states,
the typical time scales
of the intra basin and inter basin dynamics  
differ by several orders of magnitude and hence the  
separation of intrabasin and interbasin variables becomes
meaningful.
 
In the $6N$-dimensional configuration space, 
the partition function $Z$ for a  
system of $N$ rigid molecules can be written as:
\begin{equation}
Z = \frac{\Lambda_x \Lambda_y \Lambda_z}
{\lambda^{3N}} \int d{\bf q}^N \exp (- V({\bf q}^N)/k_B T),
\end{equation}
where ${\bf q}^N$ denotes the positions and orientations of 
the molecules, $ V({\bf q}^N)$ is the potential energy , 
$I_\mu$, with $\mu=x,y,z$, are the principal moments of 
inertia of the molecule, 
${ \Lambda}_\mu \equiv (2\pi I_\mu k_B T)^{1/2}/ h$,
and $\lambda\equiv h (2\pi m k_B T)^{-1/2}$ is the de Broglie wavelength. 

Let $\Omega(E_{IS})$ denote the number of minima with energy
$E_{IS}$, and $f(T, E_{IS})$ the average free energy of a basin with basin 
depth $E_{IS} $. $f(T, E_{IS})$, which takes into account both
the kinetic energy of the system and the local structure of the basin with
energy $E_{IS}$, is defined by:
\begin{eqnarray}
f(T, E_{IS})&\equiv& -k_BT \ln \left [\frac{\Lambda_x \Lambda_y \Lambda_z}
{\lambda^{3N}} \frac{1}{\Omega(E_{IS})}\right . \times \nonumber\\
&&\left .\sum_{\mbox{\scriptsize basins}}
\int_{R_{\mbox{\scriptsize basin}}} d{\bf q}^N
\exp (-(V-E_{IS})/k_BT)  \right ], 
\end{eqnarray}
where $R_{basin}$ is the configuration volume associated 
with the specific basin. 
The partition function can then be rewritten as  a sum over all 
basins in configurational space, i.e.
\begin{eqnarray}
Z = \sum_{E_{IS}} \Omega(E_{IS})
\exp\left({-\frac{E_{IS}+f(T,E_{IS})}{k_BT}}\right) \nonumber  \\
= \sum_{E_{IS}} \exp\left({-\frac{-T S_{c}(E_{IS})+
E_{IS}+f(T,E_{IS})}{k_BT}}\right)
\label{eq:zpoefe}
\end{eqnarray}
where the configurational entropy $S_{c} (E_{IS})$ has been defined as 
\begin{equation}
 S_{c} (E_{IS}) \equiv k_B \ln\left [\Omega
(E_{IS})\right ].
\end{equation}

In the thermodynamic limit, the free energy of the liquid can be
calculated using
\begin{equation}
F\left [ e_{IS}(T)\right ]=e_{IS}(T)+
f\left [ T,e_{IS} (T)\right ]-T S_c\left [ e_{IS}(T)\right ], 
\label{free_energy}
\end{equation}
where $e_{IS}(T)$, the average value of the IS energy 
at temperature $T$, is the solution of the 
saddle point equation
\begin{equation}
1+\frac{\partial f}{\partial E_{IS}}
-T\frac{\partial S_c}{\partial E_{IS}}=0.
\label{eq:saddle}
\end{equation}
The liquid free energy expression Eq.~(\ref{free_energy})
has a clear interpretation. 
The first term in Eq.~(\ref{free_energy}) takes into 
account the average energy of the PES minimum visited, 
the second term describes the volume of the corresponding 
basin of attraction and the kinetic energy, 
and the third term is a measure of the multiplicity of the basin.

It can be rigorously shown~\cite{heuer,jpcm00,sastry01} that, 
if the density of state $\Omega(E_{IS})$ is Gaussian, 
and if the basins have approximately the same
shape or are, to a good degree, harmonic, 
the important relation holds
\begin{equation}
e_{IS}(T)\propto \frac{1}{T}.
\label{inh_temp}
\end{equation}
On lowering $T$, basins with lower $E_{IS}$ energies   
and lower degeneracy are populated, i.e., 
both $e_{IS}$  and $S_c$ decrease with $T$.
\section{Evaluation of the configurational entropy}
\label{configurational_entropy_2}

The Eq.~(\ref{free_energy}) provides a natural starting point
for a numerical evaluation of the configurational entropy.
Indeed, the free energy $F(T,V)$ per molecule can be split in the 
usual way as a sum of an energy and an entropic contribution.
Considering Eq.~(\ref{free_energy}) we write:
\begin{eqnarray}
F(T)&=&E(T)-T S(T)\\
&=& - T S_c(T) +e_{IS}(T)+
E_v(T)-T S_v(T)\nonumber
\end{eqnarray}
where the index $v$ indicates the vibrational 
quantities (intra-basin components). 
In order to evaluate these quantities we calculate
the basin free energy as the free energy 
of $6N-3$ independent harmonic oscillators~\cite{scala00} 
plus a contribution that takes into account the basin anharmonicities.
Then we can write
\begin{eqnarray}
E(T)&=&\left (6- \frac{3}{N}\right )\frac {k_B T}{2}+e_{IS}(T)+U_{\rm anh}(T),
\label{e_totale}\\
S(T)&=&S_v(T)+S_c(T)\nonumber\\
&=&S_{\rm harm}(T)+S_{\rm anh}(T)+S_c(T),
\label{Sot}
\end{eqnarray}
and
\begin{equation}
S_{\rm harm}=\left (6-\frac{3}{N}\right)-\frac{1}{N}
\sum_{n=1}^{6N-3} \ln \left[\frac{\hbar\;
\omega_n(T)}{k_BT}\right],
\label{S_harmonic}
\end{equation}
where the frequencies $\omega_n$ are the square root of the 
eigenvalues of the Hessian matrix calculated 
in the inherent structures.

Thus, the total entropy is the sum of two contributions:
$S_c(T)$ which accounts for the multiplicity of 
basins of depth  $e_{IS}(T)$, and $S_v(T)$
which accounts for the ``volume'' of the basins. 
The last equations give us, in a very transparent
way, the physical meaning of the partition of the PES;
moreover, they provide us a very efficient way to calculate
the configurational entropy as a difference between the
total energy of the system and the vibrational entropy.

The total entropy $S$ can be evaluated via
thermodynamic integration, starting from a known reference
point.  Every variation of total entropy can be generally written as
the sum of variation along isochores and isotherms in the form:
\begin{equation}
\Delta S=\Delta S_V + \Delta S_T.
\end{equation}
Then the change of entropy along an isochore
between two temperatures $\bar{T}$ and $T$ is 
\begin{eqnarray}
\label{eq:intsv}
\Delta S_V&=&S(V,T)-S(V,\bar{T})\\
&=&\int_{\bar{T}}^T \frac{dT'}{T'} \; c_v (T') \nonumber
\end{eqnarray}            
and the change along an isotherm between two volumes
$\bar{V}$ and $V$ is
\begin{eqnarray}
\Delta S_T&=&S(V,T)-S(\bar{V},T)\\
&=&\frac{1}{T}\left [ E(V,T)-E(\bar{V},T)+
\int_{\bar{V}}^{V} d \bar{V} P(\bar{V},T) \right ]\nonumber.
\end{eqnarray}            
In the present case, to evaluate the total entropy of the liquid 
we start from the known expression of the ideal gas of LW molecules, 
reviewed in Appendix A. To evaluate the basin free energy 
$f\left [ T,e_{IS} (T)\right ]$, we select as reference
point the  free energy of $(6N-3)$ independent
harmonic oscillators  (whose distribution of
frequencies can be calculated evaluating the eigenvalues
of the Hessian matrix evaluated in the IS structure)
and add corrections to take into account 
the basin anharmonicities.The harmonic contribution 
to the entropy is given by Eq.~(\ref{S_harmonic}).

Assuming that the anharmonic contribution is
independent from the basin depth, 
the anharmonic corrections to the entropy at $T$ 
can be calculated integrating the quantity $dU_{\rm anh}/T$,
where $U_{\rm anh}$ is implicitly defined in Eq.~(\ref{e_totale}),
from $T=0$ to $T$ (see Eq.~(\ref{eq:intsv})).
\section{NUMERICAL DETAILS}
\label{numerics}
The LW model is defined by the spherical potential  
\begin{equation}
V(r)= 4 \epsilon\,\left [ \left ({{\sigma}\over {r}}\right )^{12} -\left
({{\sigma}\over
{r}}\right )^6\right ]
+ \lambda_1 + \lambda_2\, r, 
\end{equation}
with $\epsilon=5.276$ $kJ/mol$, $\sigma=0.483$ nm
$\lambda_1 = 0.461$ $kJ / mol$ and $\lambda_2=-0.313$ $kJ /( mol~nm)$.
The parameters of the potential are selected to reproduce
some bulk properties of the OTP molecule~\cite{lewis}
such as the temperature dependence of the diffusion 
coefficient and the structure. The values of $\lambda_1$ and  
$\lambda_2$ are selected in such a way the potential and 
its first derivative are zero at $r_c=1.2616$ $nm$.  
Such a potential is characterized 
by a minimum at $r=0.542$ $nm$ of depth $-4.985$ $kJ/mol$.
The integration time step is $0.01$ ps. The shake algorithm 
is implemented to account for the molecular constraints. 

We study a (N,V,E) system composed by $N=343$ molecules
($1029$ LJ interaction sites) at 5 different densities 
(see Table~\ref{numerics1:table}) 
for several temperatures at each density 
(Table~\ref{numerics2:table}). 
The total simulation time is quite long, exceeding 10 $\mu s$.
We take care to check the thermalization of the system at the
lowest temperatures; the production run follows a thermalization run 
whose length is comparable to the relaxation time at the considered
thermodynamical point.
We are able to thermalize the system at
temperatures at which the diffusion constant 
reaches values very low, of order 10$^{-10}$ cm$^2$/s,
10$^4$ smaller then at ambient temperature.

Two additional simulations are  performed to connect 
the range of densities and temperature studied with
the ideal gas reference point. 
The system at density $\rho_4$ is simulated for temperatures ranging
from 250 to 5000 K  to evaluate the $T$ dependence of 
the potential energy.
A second set of simulations at constant $T$ ($T=5000$ K) 
in the volume range $10^2- 10^5$ nm$^3$ is performed 
to calculate the {\em excess} pressure ( i.e. the pressure beyond the 
ideal gas contribution).

To calculate the inherent structures visited in equilibrium
we perform conjugate gradient minimizations 
to locate the closest local minima on the PES. We use a
tolerance of $10^{-15}$ kJ/mol in the total energy for the minimization.
For each thermodynamical point we minimize at least $100$ 
configurations and we diagonalize the Hessian matrix 
of at least $50$ configurations 
to calculate the density of states. The Hessian is
calculated choosing  for each molecule  the center of mass 
and the angles associated with rotations around the three 
principal inertia axis as coordinates.
\section{RESULTS}
\label{results}
\subsection{Dependence of the total entropy on $T$ and $\rho$}
\label{thermo_integration}
\begin{figure}
\hbox to\hsize{\epsfxsize=1.0\hsize\hfil\epsfbox{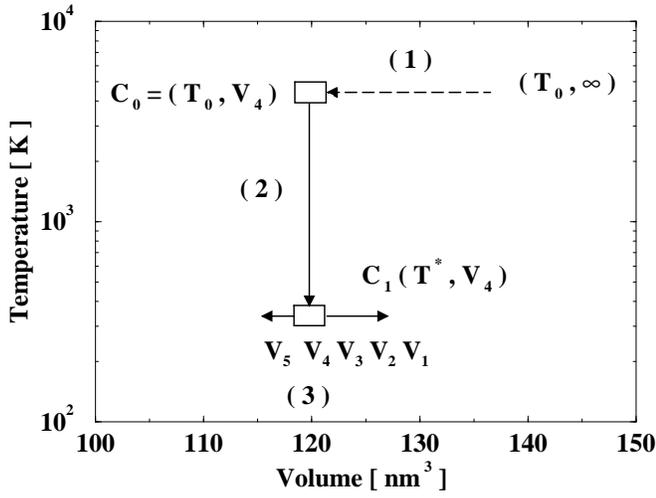}
\hfil}
\caption{Thermodynamic integration paths used to calculate 
the total entropy at the thermodynamical points of interest
starting from the ideal ---non interacting-- gas state.
Details are given in the text.
}
\label{fig:thermopath}
\end{figure}

To estimate the total entropy for the model 
we proceed in three steps as shown in Fig.~\ref{fig:thermopath}.
The thermodynamic path has been chosen to avoid the liquid-gas
first order line.

\begin{figure}
\hbox to\hsize{\epsfxsize=1.0\hsize\hfil\epsfbox{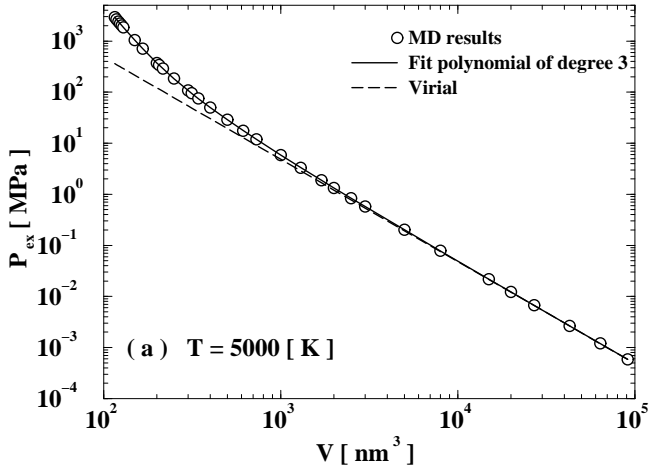}
\hfil}
\hbox to\hsize{\epsfxsize=1.0\hsize\hfil\epsfbox{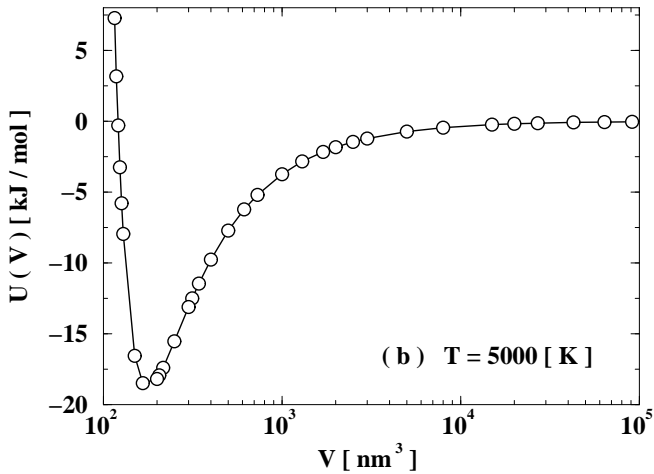}
\hfil}
\caption{(a) Excess pressure at $T=5000$ K as a function of volume. 
The open circles are the MD results. 
The dashed line is the the first term of the 
virial expansion to the excess pressure; the solid line
is a third order polynomial fit to the entire set of data. 
(b) Potential energy at $T=5000$ K as a function of volume.
}
\label{fig:fit_Pex}
\end{figure}

{\em (1)} Integration along the isotherm $T_0$ = 5000 K
from ($T_0$,$V=\infty$) (perfect gas) to ($T_0$, V$_4$=118.35 nm$^3$), 
corresponding to point $C_0$ in Fig.~\ref{fig:thermopath}.
The ideal gas contribution to the total entropy 
is discussed in Appendix A.
The entropy at $C_0$ can be calculated as
\begin{eqnarray}
S(T_0,V_4)-S_{id}(T_0,V_4)=&& \nonumber\\
\frac{U(T_0,V_4)}{T_0}+\int_\infty^{V_4}\frac{dV}{T_0}P_{ex}(V,T_0),&&
\end{eqnarray}
where $P_{ex}$ is the pressure that exceeds the pressure of
the ideal gas, i.e. the contribution to the pressure due to the
interaction potential and $U$ is the system potential energy.
The values of the pressure $P_{ex}(T=T_0,V,N=343)$ as a function of $V$
are reported in Fig.\ref{fig:fit_Pex} (a). 
$P_{ex}(T=T_0,V,N=343)$ has been fit using the virial
expansion
\begin{equation}
P_{ex}(T=T_0,V,N=343)=\sum_{k=1}^4 a_k V^{-(k+1)}.
\end{equation}
The $a_k$ values are reported in Table~\ref{fit_press:tab}, 
from which we estimate the  first virial coefficient
at $T_0$
\begin{equation}
B_2(T_0)=a_1/(k_B T_0 N^2)=0.596 \;{\rm nm^3}.
\end{equation}
In Fig.\ref{fig:fit_Pex} (b) we plot the potential energy as a function of
volume along the $T=T_0$ isotherm. 

The total entropy at the reference point $C_0$
is $S(C_0)=$ 294.8 J/( mol K ), resulting from the sum
of three contributions
\begin{eqnarray}
&&S_{id}(C_0)=339.03 \;J/(mol K), \\
&&\int_{\infty}^{V_4} \frac{dV}{T_0}P_{ex}(V,T_0)=-44.9 \;J / (mol K),
\end{eqnarray}
and
\begin{eqnarray}
&&\frac{U(C_0)}{T_0}= 0.64 \;J / (mol K)
\end{eqnarray}

\begin{figure}
\hbox to\hsize{\epsfxsize=1.0\hsize\hfil\epsfbox{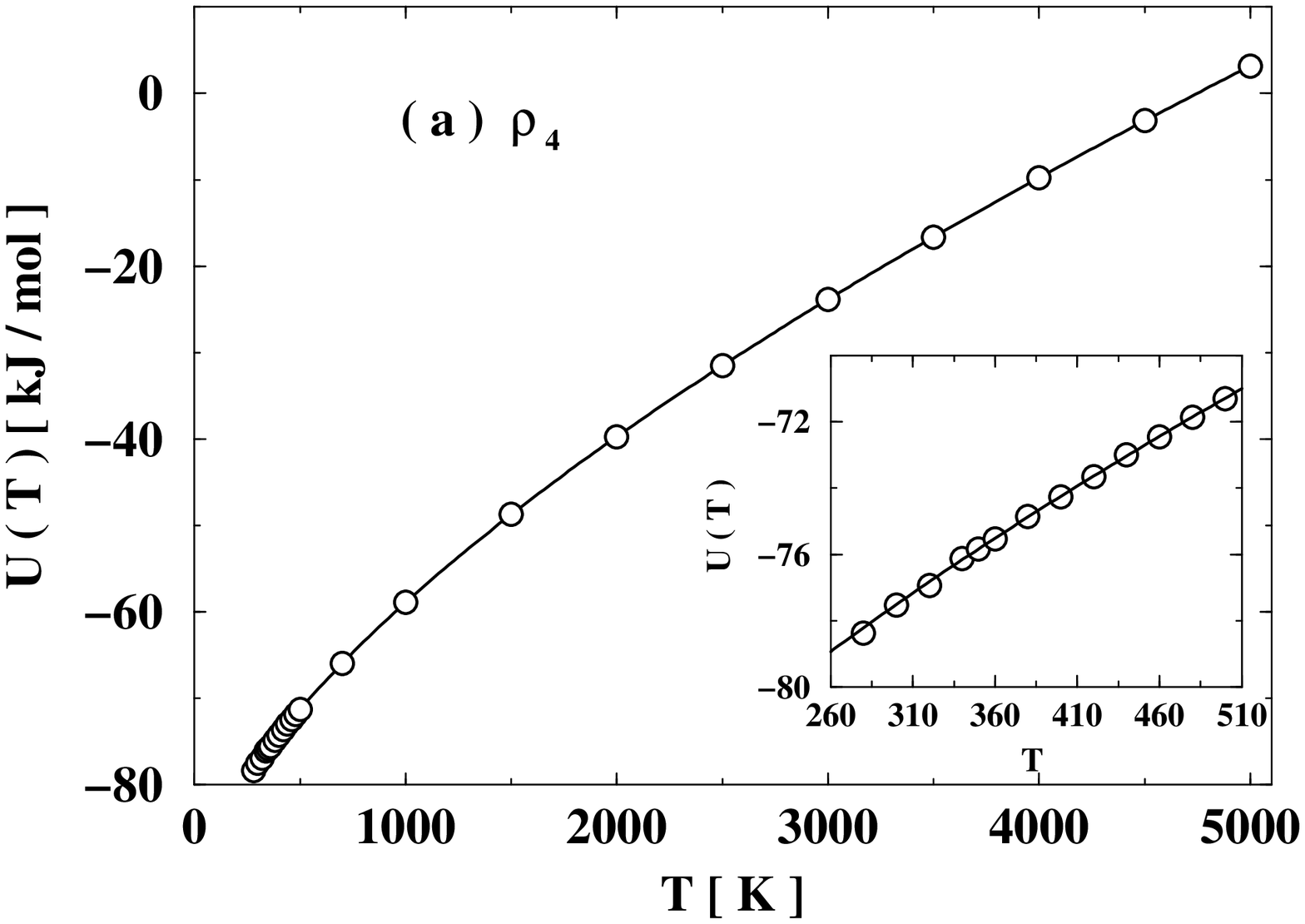}
\hfil}
\hbox to\hsize{\epsfxsize=1.0\hsize\hfil\epsfbox{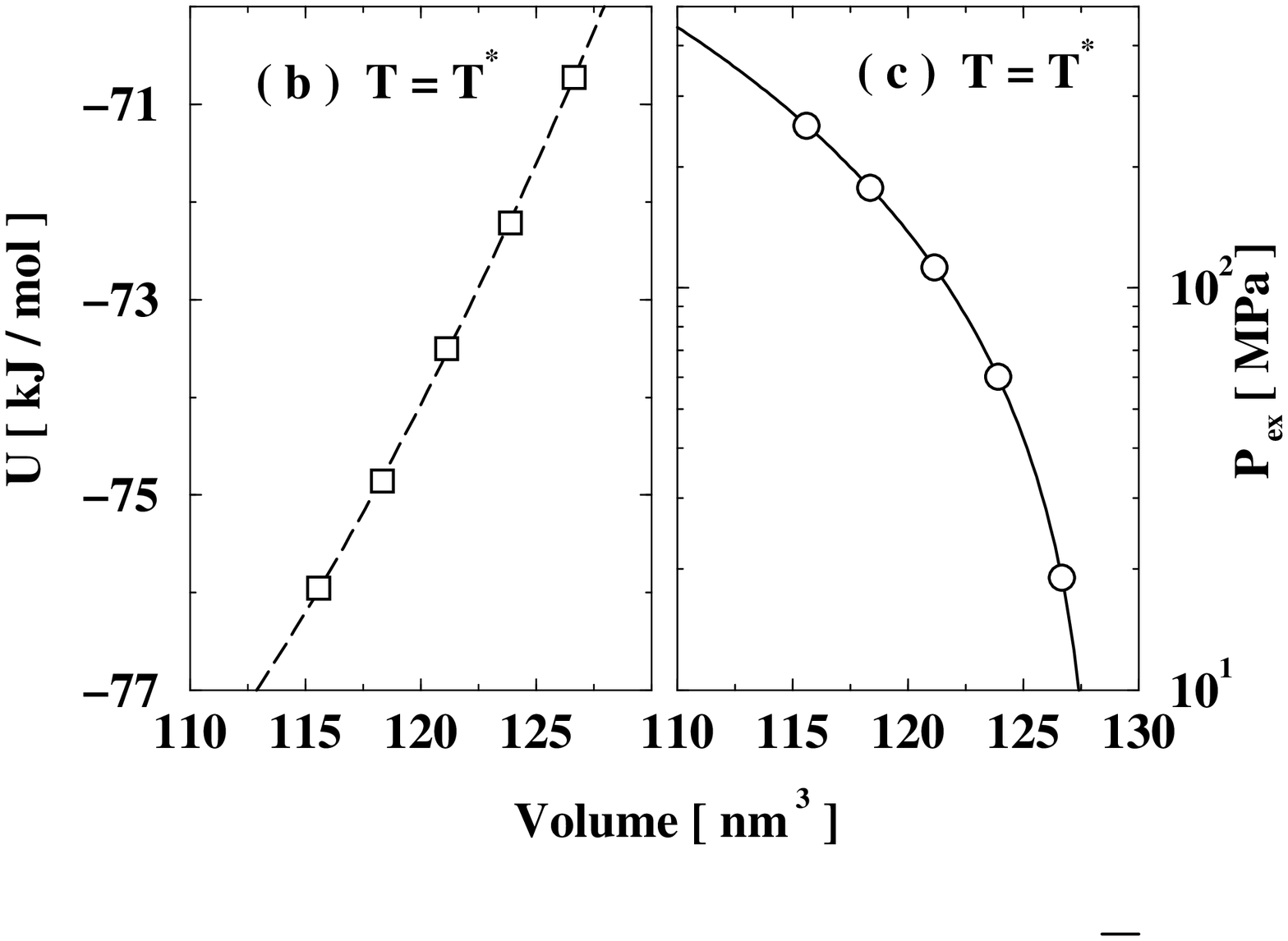}
\hfil}
\caption{
(a) Integration Step 2. 
Potential energy (open circles) at the density $\rho_4$
in the entire temperature range considered; the solid line 
is the fit of the data to Eq.~(23). The inset shows the
lowest temperatures region in order to stress 
the accuracy of the fit.
(b) and (c) Integration Step 3. Potential energy (b) and pressure (c).
}
\label{fig:ene_rho4}
\end{figure}

{\em (2)} Integration along the isochore $V=V_4$ from $T_0$ to 
$T^*=380$ K, corresponding to the point $C_1$ in Fig.\ref{fig:thermopath}.
To evaluate the entropy along this isochore we use
\begin{eqnarray}
S(T^*,V_4)& =& S(T_0,V_4) + 3 R \; \log(T^*/T_0) \nonumber \\
&+&\int_{T_0}^{T^*}\frac{dT}{T}\frac{\partial U(V_4,T)}{\partial T}.
\label{eq:c2}
\end{eqnarray}
Fig.~\ref{fig:ene_rho4} (a) shows the potential energy for the
$V=V_4$ isochore. To calculate the integral in Eq.~(\ref{eq:c2}), 
we fit the potential energy using the functional form which
best interpolates the calculated points
\begin{equation}
U(V_4,T)=u_0+u_1 T^{3/5}+u_2 T,
\label{eq:taraz_plus_lin}
\end{equation}
obtaining the values
$u_0=-94.405 , u_1=0.533 , u_2 = 0.00183$ (energy in $kJ / mol$).

\begin{figure}
\hbox to\hsize{\epsfxsize=1.0\hsize\hfil\epsfbox{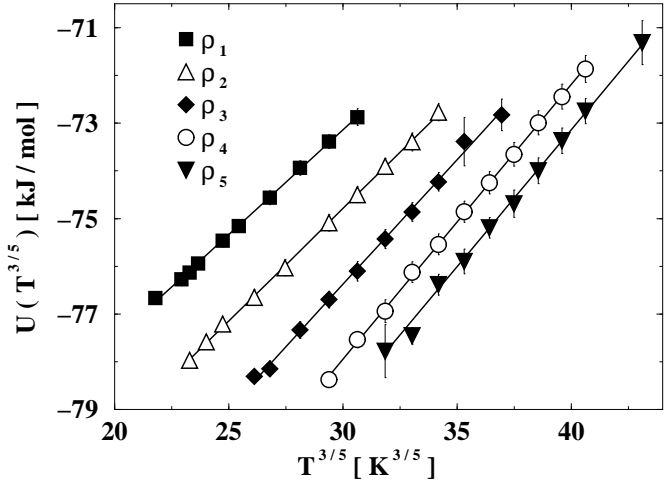}
\hfil}
\caption{Potential energies at the different densities 
as a function of $T^{3/5}$. The straight solid lines 
show the validity of the Rosenfeld-Tarazona law Eq.~(29).}
\label{fig:potential}
\end{figure}

The total entropy at the reference point $C_1$
is $S(C_1)=$ 191.8 J/( mol K ), resulting from the sum
of three contributions:
\begin{eqnarray}
&&S(C_0)= 308.6 \;J/(mol K), \\
&& 3 R \; \log(T/5000)   = -64.3\;J / (mol K),
\end{eqnarray}
and
\begin{eqnarray}
&& \int_{T_0}^{T^*}\frac{dT}{T}\frac{\partial U(V_4,T)}{\partial
T}=-52.5 \;J / (mol K).
\end{eqnarray}
{\em (3)} Integration along the isotherm $T^*$ from $V_4$ to a ``generic'' $V$.
To determine the total entropy difference for all studied densities
we calculate
\begin{eqnarray}
S(T^*,V)-S(T^*,V_4)&=&S_{id}(T^*,V)-S_{id}(T^*,V_4)\nonumber\\
&+&\frac{1}{T^*}\left [U(T^*,V)-U(T^*,V_4)\right ]\nonumber\\
&+&\int_{V_4}^{V}\frac{dV'}{T^*}P_{ex}(T^*,V').
\end{eqnarray}
Figs.~\ref{fig:ene_rho4} (b) and (c) show respectively the potential
energy and the excess pressure as a function of volume at $T=T^*$. 
For convenience we fit $P_{ex}$ with a third order polynomial
\begin{equation}
P_{ex}(T^*,V)=\sum_{k=1}^{4} p^*_k V^{k-1},
\end{equation}
where the values of the coefficients $p^*_k$ 
are given in Table~\ref{fit_press:tab}.
The resulting total entropy at $T^*$ for all studied densities is
reported in Table~\ref{total_ref:tab}. These values are used as 
reference entropies for the $T$ dependence of $S$.
%
For each of the studied isochores, we calculate the $T$-dependence
of the total entropy according to Eq.~(\ref{eq:c2}). In this low
$T$-range, the potential energy is very well represented by the
Rosenfeld-Tarazona law~\cite{tarazona}
\begin{equation}
U(V,T)=U_0(V)+\alpha(V) \; T^{3/5}
\end{equation}
consistent with what was found for LJ systems.
In Fig.~\ref{fig:potential} we show the temperature dependence of the 
potential energy at all densities.
The best-fit  $U_0(V)$  and $\alpha(V)$ values are 
reported in Table~\ref{energy_fit:tab}.

\begin{figure}
\hbox to\hsize{\epsfxsize=1.0\hsize\hfil\epsfbox{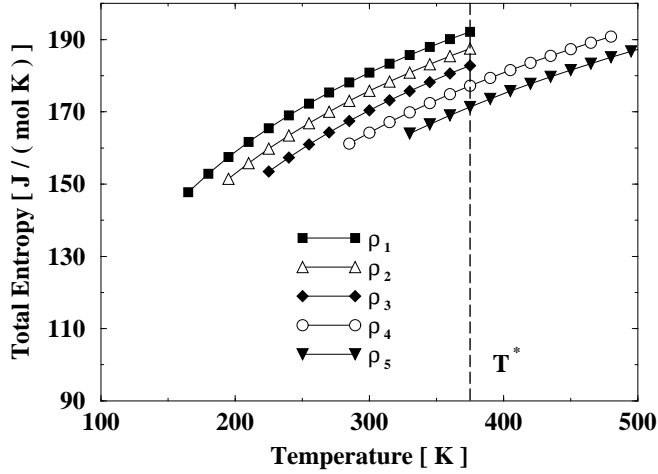}
\hfil}
\caption{Temperature dependence of the total entropy 
as calculated by thermodynamic integration from the ideal gas
reference state. Only points in the temperature range 
where MD measurements
have been performed are shown. The reference temperature $T^*=380$ K is
also shown (dashed line).}
\label{fig:s_totale}
\end{figure}

The calculated total entropies at each considered density
are plotted in Fig.~\ref{fig:s_totale}.
\subsection{Dependence of the inherent structure 
energies on $T$ and $\rho$}
\label{inherent structures}
\begin{figure}
\hbox to\hsize{\epsfxsize=1.0\hsize\hfil\epsfbox{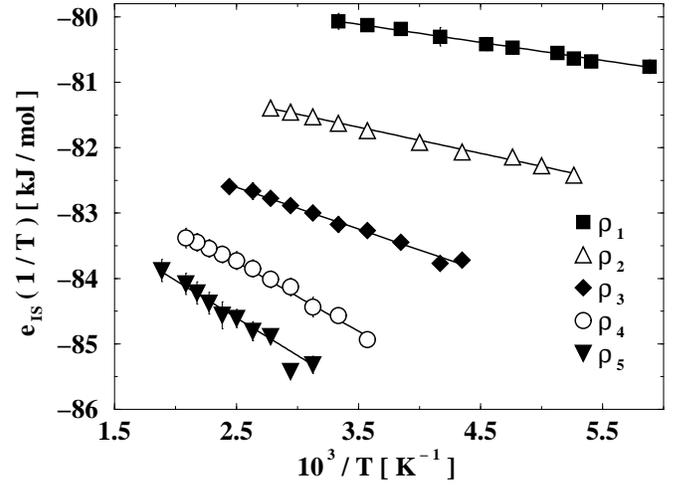}
\hfil}
\caption{Energies of the inherent structures 
at the different densities as a function of
$1/T$. The straight lines confirm the validity of Eq.~(30)
in the entire temperature range considered.}
\label{fig:inherent}
\end{figure}

In Fig.~\ref{fig:inherent} we show the temperature dependence of the
energy of the calculated inherent structures together with a fit 
(according to Eq.~(\ref{inh_temp})) in the form
\begin{equation}
e_{IS}(V,T)=A(V) + \frac{B(V)}{T}
\label{inherent_fit}
\end{equation}
The values of the fitting coefficients $A(V)$ and $B(V)$
are reported in Table~\ref{energy_fit:tab}.
On lowering temperature the system populates minima of lower and lower
energy. It is worth noting that, in contrast to the case of the actual
potential energy, the slope of these curves varies strongly with
densities. 

\begin{figure}
\hbox to\hsize{\epsfxsize=1.0\hsize\hfil\epsfbox{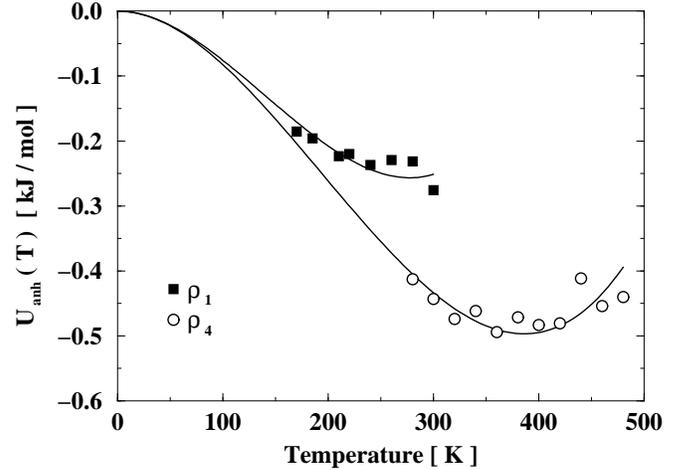}
\hfil}
\caption{Anharmonic contributions to the energies, 
at the two indicated densities, together
with the appropriate cubic fit Eq.~(31). 
This contribution is integrated to directly 
calculate the anharmonic contribution
to the vibrational entropy.}
\label{fig:anharmonic}
\end{figure}

From the $T$ and $V$ dependence of $e_{IS}$ the
anharmonic potential energy can be calculated.  
Fig.~\ref{fig:anharmonic} shows $U_{\rm anh}(T)$
for two densities (symbols).  We also show a 
cubic extrapolation (solid lines) in the form of
\begin{equation}
U_{\rm anh}(T)=c_2 T^2+c_3 T^3.
\label{anh_energy}
\end{equation}
As shown in Fig.~\ref{fig:anharmonic}, the anharmonic 
contribution is rather small, in agreement with previous 
findings for the LJ model. For this reason, the
low signal to noise level does not allow a well-defined 
characterization of the $c_2$ and $c_3$  values. 
To decrease the number of free parameters, we 
consider $c_2$ to be volume independent, and we
fit simultaneously, according to Eq.~(\ref{anh_energy}),  
$c_2$ and the $V$ dependence of $c_3$. 
As we will show in the following, the anharmonic contribution
to the entropy is much smaller than the
harmonic one and hence the choice of $c_2$ and $c_3$ 
does not affect significantly the resulting configurational
entropy estimate.
\subsection{Density of states and vibrational harmonic entropy}
\label{density_of_state}
\begin{figure}
\hbox to\hsize{\epsfxsize=1.0\hsize\hfil\epsfbox{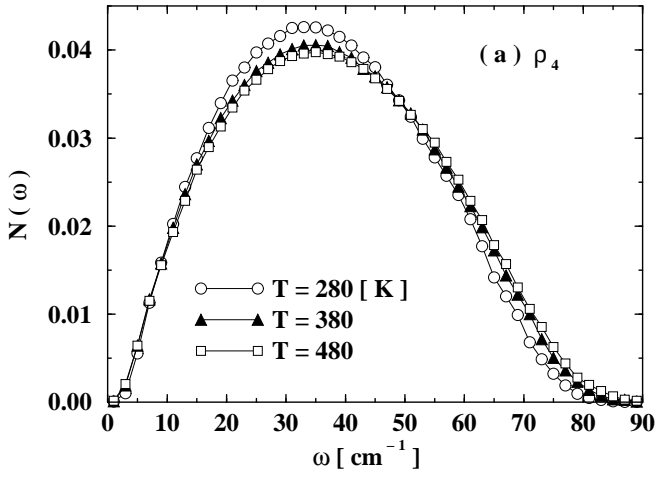}
\hfil}
\hbox to\hsize{\epsfxsize=1.0\hsize\hfil\epsfbox{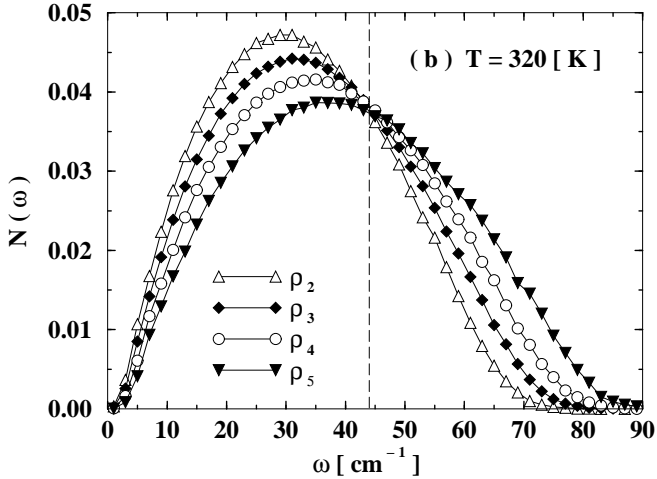}
\hfil}
\caption{(a) Density of states at fixed density $\rho_4$ at 
the three indicated temperatures. This quantity
is the histogram of the square root of the eigenvalues of the Hessian
calculated for the inherent structures.
(b) Density dependence of the density of state at fixed 
temperature $T=320$ K. The dashed line indicates 
the isosbestic frequency $\omega^*\approx 44$ cm$^{-1}$ at
which all the curves intersect. The relevance of this feature
is discussed in the text.}
\label{fig:density_of_state_fig}
\end{figure}

In this section we study the shape of the basins
by investigating the properties of the density of states
and we calculate the vibrational harmonic entropy.
In Figs.~\ref{fig:density_of_state_fig} (a) and (b) we show the 
temperature and density dependence of the density of state, namely 
the histogram of the square root of the 
eigenvalues of the Hessian calculated for the 
inherent structures.
The distribution is characterized by only one peak,
not showing any clear separation between translational and
rotational dynamics; the width of the distribution increases
on increasing temperature.
The position of the maximum is found to be to a good extent
independent of temperature; at variance it increases 
with density as the width does.
These features show that the LW PES basins have shapes 
that are function of the energy depth and of the density.

It is worth noticing one particular feature of
Fig.~\ref{fig:density_of_state_fig} (b);
all the curves cross at a value of the 
frequency $\omega^*\approx 44$
cm$^{-1}$. The presence of  this isosbestic frequency
(in analogy with the well-know  isosbestic frequency 
observed in the Raman spectrum of water\cite{darrigo})
supports the possibility that a two-state 
model~\cite{austenrecent} may provide a reasonable description of the
change of the density of states with temperature and, correspondingly, 
of the change of the density of states with the basin depth.

\begin{figure}
\hbox to\hsize{\epsfxsize=1.0\hsize\hfil\epsfbox{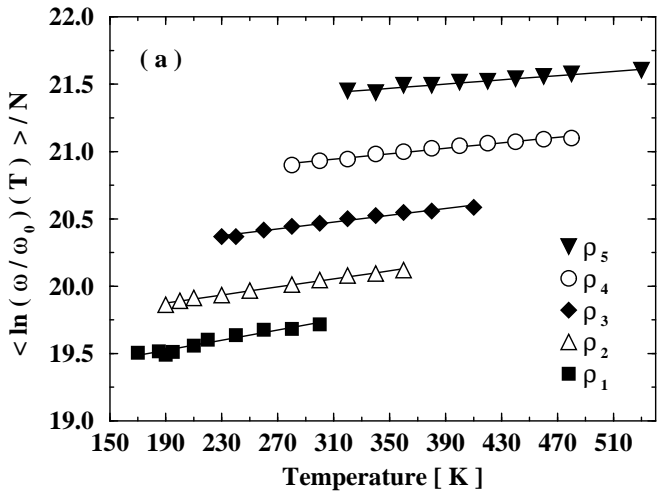}
\hfil}
\hbox to\hsize{\epsfxsize=1.0\hsize\hfil\epsfbox{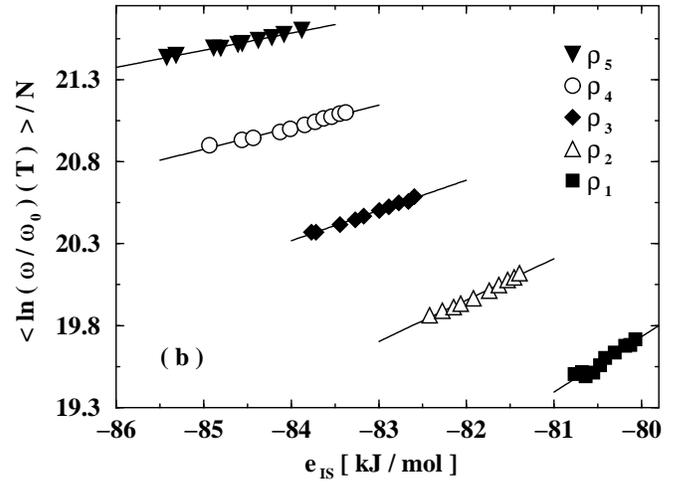}
\hfil}
\caption{
(a) Temperature dependence of the average basin curvatures
$N^{-1} \sum_{k=1}^{6N-3}\log(\omega_k/\omega_o)$; this quantity,
being a sum of logarithms, is very sensitive to the spectrum tails.
$\omega_o=1$ cm$^{-1}$ sets the frequency scale.
(b) Relation between the energy of the inherent 
structures and the average basin curvatures. 
The straight lines confirm the correlation between shape
and depth of the inherent structures accessed by the system.
}
\label{fig:log_omega_fig}
\end{figure}

In Fig.~\ref{fig:log_omega_fig} (a) and (b) we plot 
the quantity $N^{-1} \sum_{k=1}^{6N-3}\log(\omega_k/\omega_o)$ 
as a function of $T$ and of the $e_{IS}$ respectively. 
The scale frequency $\omega_o$ is chosen as $1$ cm$^{-1}$.
This quantity is an indicator of the average curvature of the 
basins and, being a sum of logarithms, is 
very sensitive to the spectrum tails. 
As shown in Fig.~\ref{fig:log_omega_fig} (a)
$N^{-1} \sum_{k=1}^{6N-3} \log(\omega_k/\omega_o)$  increases 
with temperature along isochores
and increases with density along isotherms. 

As noted previously for the LJ\cite{st01,sastry01} 
and for the simple-point charge extended (SPC/E) model
for water~\cite{starr01}, the dependence of 
$N^{-1} \sum_{k=1}^{6N-3}\log(\omega_k/\omega_o)$ from 
$e_{IS}$ can be well approximated by a linear dependence, i.e.
\begin{equation}
\frac{1}{N}\sum_{k=1}^{6N-3}\ln \left[\frac{\hbar\;
\omega_n(T)}{k_B T_o}\right]=a(V)+b(V) \; e_{IS}(T),
\label{log_omega_vs_eis}
\end{equation}
where $T_o$ defines the $T$ scale ($T_o=1 K$). 
This dependence indicates that deeper and deeper basins have
larger and larger volumes (their average frequency being
smaller). The fact that basins of different depths have 
different volumes introduces an important contribution
to Eq.~(\ref{eq:saddle}) since the term 
$\partial f / \partial e_{IS}$ is different from zero.
The implication of this non-zero contribution has been
discussed recently in Refs.~\cite{sastry01,st01,austen01}.

\begin{figure}
\hbox to\hsize{\epsfxsize=1.0\hsize\hfil\epsfbox{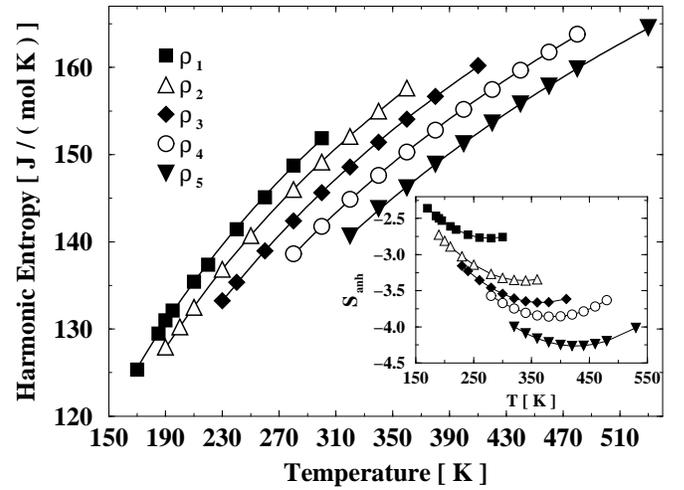}
\hfil}
\caption{Main panel: Harmonic contribution to the vibrational 
entropy as calculated from the eigenvalues of the Hessian
for the inherent structures.
Inset: Anharmonic contribution to the vibrational entropy 
as calculated by integration of the anharmonic contribution
to the potential energy, as discussed in the text.}
\label{fig:s_harmonic_fig}
\end{figure}

In Fig.~\ref{fig:s_harmonic_fig} we 
show the harmonic contribution to 
the entropy as calculated from Eq.~(\ref{S_harmonic}).
This contribution is obviously increasing with temperature
and along isotherms increases decreasing density.
The lines are interpolations of the data
using the fits of Fig.~\ref{fig:log_omega_fig}. 
\subsection{Vibrational anharmonic entropy}
\label{anh_entropy}
Integration of the anharmonic energy $U_{\rm anh}$, obtained from 
Eq.~(\ref{e_totale}) according to Eq.~(\ref{eq:intsv}),
gives directly the anharmonic contribution to the entropy. 
For the LW case, $U_{\rm anh}$ is described by 
the polynomial in $T$ of Eq.~(\ref{anh_energy}), and we obtain
\begin{equation}
S_{\rm anh}(T)=2 c_2 T + \frac{3}{2} c_3 T^2.
\label{S_anharmonic}
\end{equation}
The inset of Fig.~\ref{fig:s_harmonic_fig} 
shows the anharmonic contribution to
the vibrational entropy as calculated by integrating 
the anharmonic contribution to the potential energy. 
This contribution is negative showing that, in the
range of densities and temperatures studied, the leading anharmonic
contribution acts in the direction to decrease the volume of
the basin.
\subsection{The configurational entropy}
\label{config_entropy}
\begin{figure}
\hbox to\hsize{\epsfxsize=1.0\hsize\hfil\epsfbox{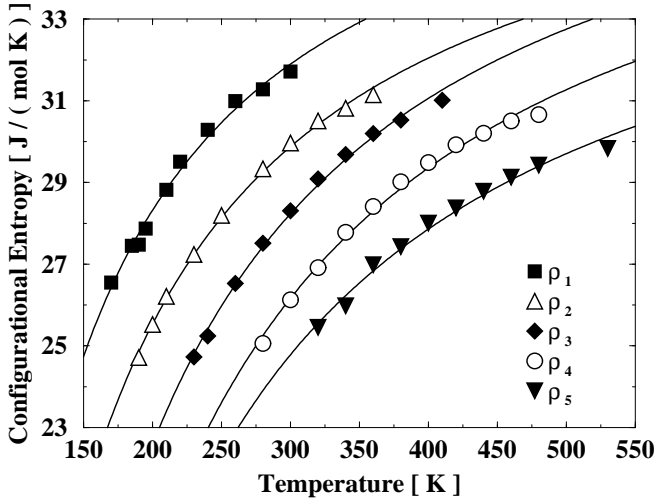}
\hfil}
\caption{Volume and temperature dependence of the 
configurational entropy $S_c$ calculated as the difference between
the total and the vibrational entropy.
Solid lines are interpolations of the calculated points to Eq.~(34).
}
\label{fig:s_config_fig}
\end{figure}

In Fig.~\ref{fig:s_config_fig} we plot the configurational 
entropy calculated subtracting the vibrational (sum of the harmonic and 
anharmonic terms) from the total entropy for the 5 studied isochores.
As expected the degeneracy of basins increases on lowering density, 
in agreement with the evidence that a glass transition may be 
induced along an isothermal path by progressively 
increasing the pressure.
Considering Eqs.~(\ref{Sot})(\ref{S_harmonic})(\ref{inherent_fit})
(\ref{log_omega_vs_eis})(\ref{S_anharmonic}),
the configurational entropy can be described in the entire 
density and temperature range considered 
by means of the functional form
\begin{eqnarray}
S_c(T)&=&S(T)-\left(6-\frac{3}{N}\right)\nonumber\\
&+&a(V)+b(V)\left[A(V)+\frac{B(V)}{T}\right]\\
&-&2 c_2 T - \frac{3}{2} c_3 T^2.\nonumber
\label{eq:sconf_tot_fit}
\end{eqnarray}
These curves are plotted in Fig.~\ref{fig:s_config_fig} as solid lines.
In the range of temperatures and density studied, 
$S_{c}/R$ per molecule varies from about 4 
to 3, a figure not very different from 
the estimated configurational entropy of
orthoterphenyl, based on an analysis of the $T$ dependence
of the measured specific heat~\cite{stillinger98,richert98}. We recall that 
the LW model represents each phenyl group as one single
interaction site and it does not account for the 
the molecule flexibility. The similar estimate of
$S_{c}$ seem to suggest that steric effects are 
dominant in controlling the configurational entropy.
\subsection{Diffusion and the Adam-Gibbs relation}
\label{diffusion}
In order to investigate the connection between the long time 
dynamics of the system and the underlying PES, we calculate the center-of-mass 
diffusion coefficient $D(T)$ from the mean-square 
displacement via the Einstein relation
\begin{equation}
D(T)=\lim_{t\rightarrow\infty}\frac{1}{6t}\langle
r^2(t)\rangle
\label{diffusion_eq}
\end{equation}
To guarantee a proper diffusive regime,
at all densities simulations are performed until the average
mean square displacement is greater than 0.1 nm$^2$ 
at the lowest temperatures and 10 nm$^2$ 
at the highest.  
The inverse of the diffusion coefficient
provides an estimate of the characteristic structural 
relaxation time of the LW model. 

\begin{figure}
\hbox to\hsize{\epsfxsize=1.0\hsize\hfil\epsfbox{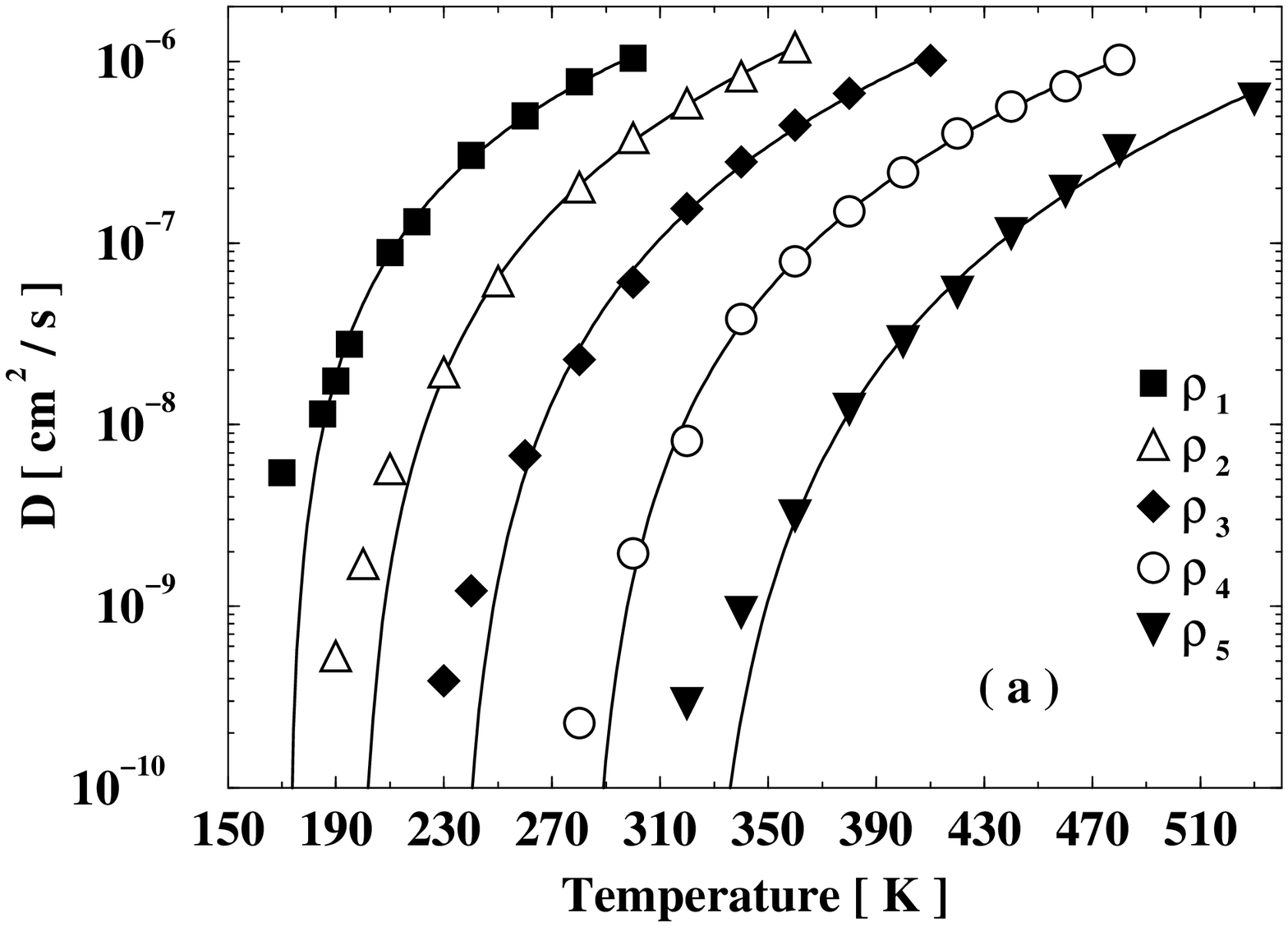}
\hfil}
\hbox to\hsize{\epsfxsize=1.0\hsize\hfil\epsfbox{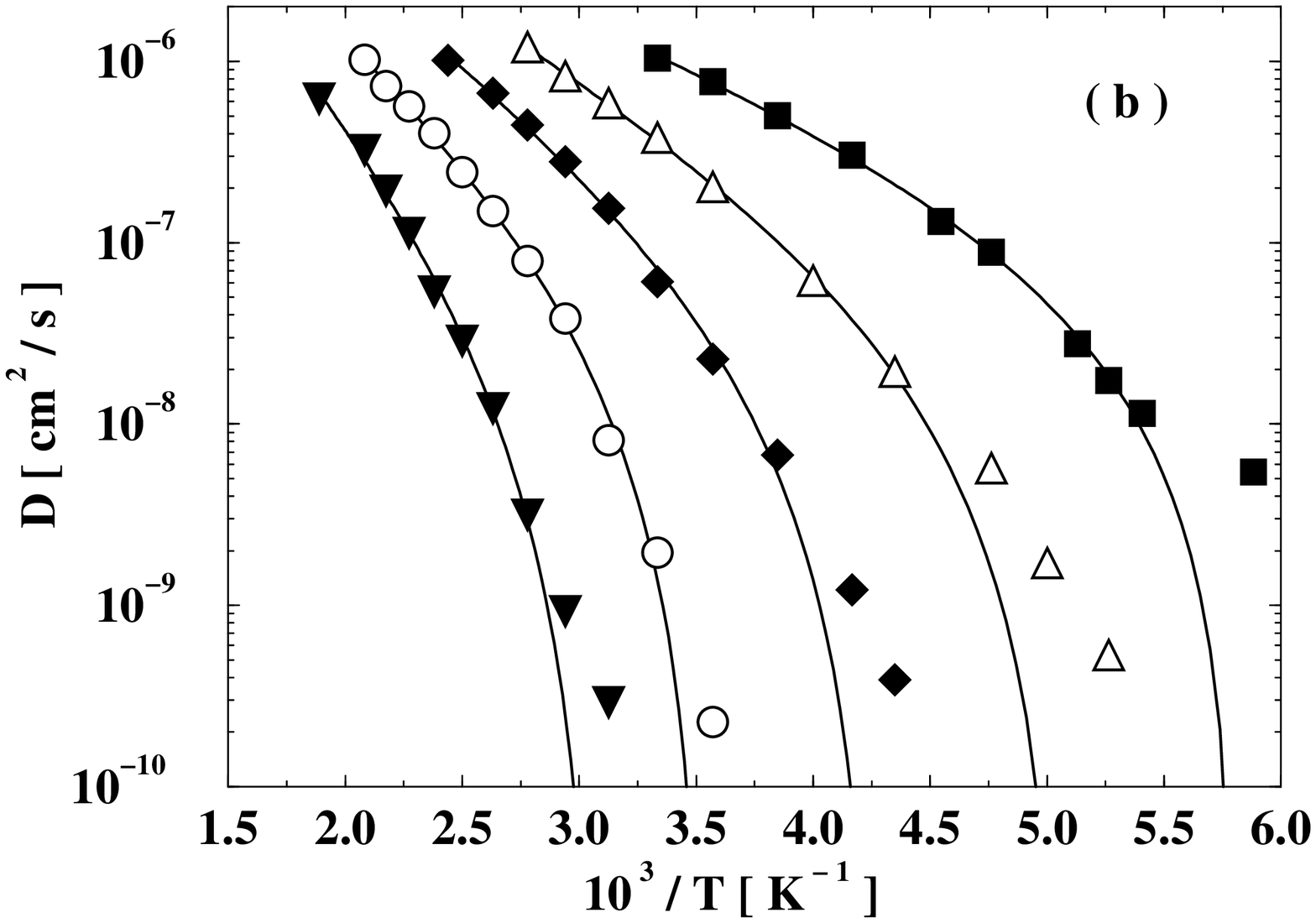}
\hfil}
\caption{Diffusion constants together with the corresponding 
power law fits (solid lines) predicted by the MCT.
The breakdown of this prediction and the crossover
to an activated dynamics is evident. See text for a discussion
of this point. (a) As a function of temperature. 
(b) As a function of the inverse 
temperature in order to stress the exponential dependence
at the lowest temperatures.}
\label{fig:diffusion_fig}
\end{figure}

The $D$ values calculated are shown in Fig.~\ref{fig:diffusion_fig}.
Fig.~\ref{fig:diffusion_fig} (a) shows the dependence on $T$ 
, while Fig.~\ref{fig:diffusion_fig} (b) shows the dependence 
on $1/T$. 
Fig.~\ref{fig:diffusion_fig} (a) also shows 
the best fits to the power law
\begin{equation}
D(T)\propto (T-T_c)^\gamma
\label{mct_diff:eq}
\end{equation}
predicted by the ideal MCT in weakly supercooled states.
The consistency of the MCT prediction for a wide range of
$D$ values confirms the analysis of 
Rinaldi {\it et al.}~\cite{rinaldi01}
where explicit ideal MCT calculations were presented and
successfully compared with the numerical results 
along one isobar.
Fig.~\ref{fig:diffusion_fig}
shows also that 
clear deviations from the ideal MCT
take place when the diffusion value becomes smaller than
$10^{-8} cm^2/s$.   The representation of $D$ as a function of
$1/T$ shown in part (b)  shows that the ideal MCT region is
followed by a $T$ region where new types of processes become
effective in controlling the molecular dynamics.
These processes, termed hopping processes,
transform the ideal MCT divergence of characteristic
times into a crossover. 
In the region of $D$ values between 
$10^{-8} cm^2/s$ and $10^{-10} cm^2/s$, limited from below by the
present numerical resources, data are consistent with 
an apparent Arrhenius dependence with parameters which could well
become $T$ dependent if studied in a larger range of $D$ values.

\begin{figure}
\hbox to\hsize{\epsfxsize=1.0\hsize\hfil\epsfbox{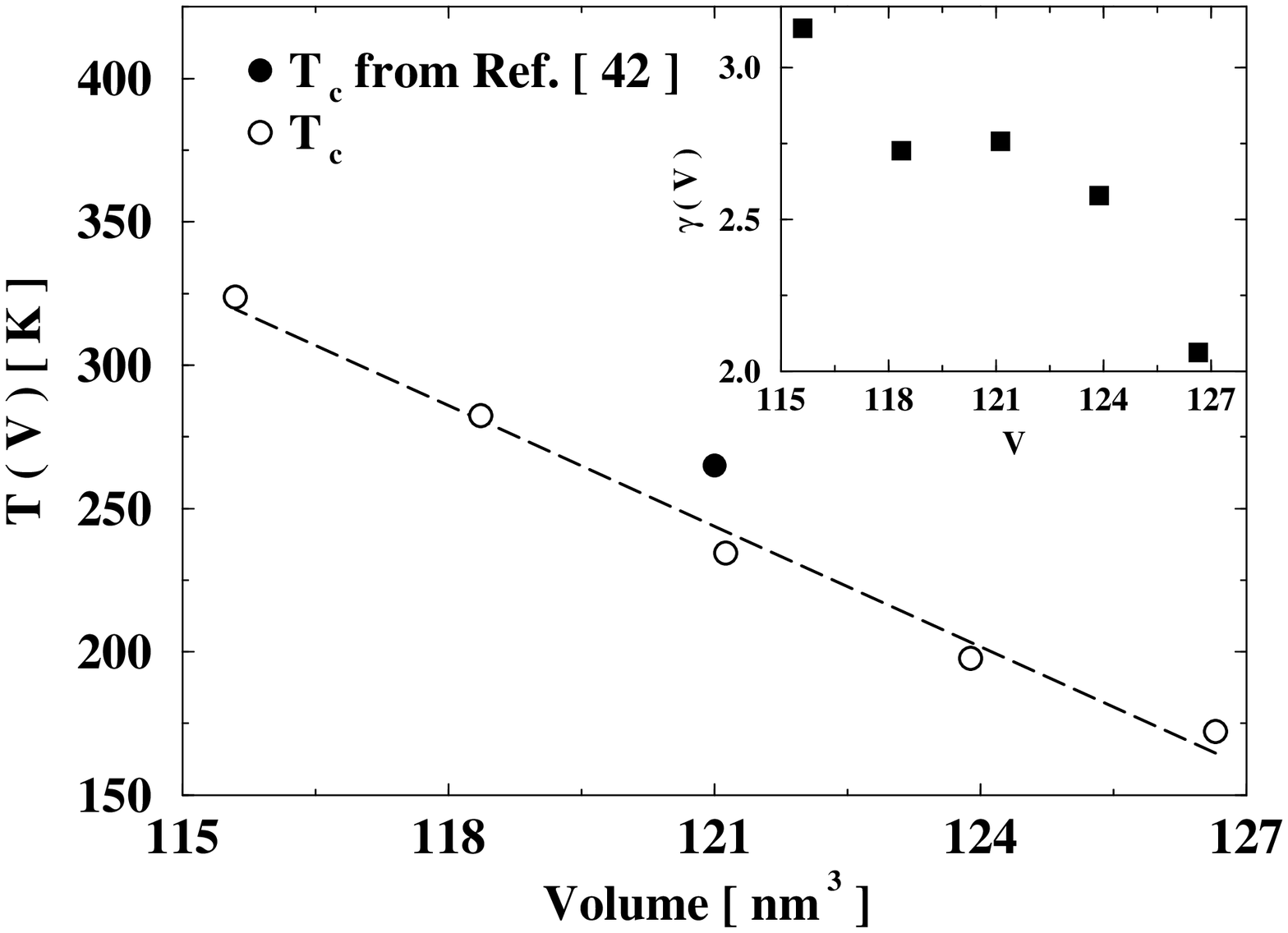}
\hfil}
\caption{MCT parameters as calculated from the 
diffusion constants. 
Main panel: Critical temperature $T_c (V)$ (open circles) together with 
the value calculated in Ref.~[42] (closed circle).
The dashed line is only a guide for the eye. 
Inset: Power law exponent $\gamma(V)$.}
\label{fig:fit_diff_fig}
\end{figure}

The ideal MCT  critical temperatures 
and $\gamma$ values, determined by the 
fit of the $D$ values to Eq.~(\ref{mct_diff:eq}), 
as a function of density  are shown 
in Fig.~\ref{fig:fit_diff_fig}.   
The density dependence of $T_c$ is 
almost linear. The exponent $\gamma$ seems 
to increase on increasing density,
but the noise does not allow us to rule out the possibility
of a constant value. The filled circle indicates the value of the critical
temperature $T_c=265$ K determined from 
an isobaric run in Ref.~\cite{rinaldi01}.

\begin{figure}
\hbox to\hsize{\epsfxsize=1.0\hsize\hfil\epsfbox{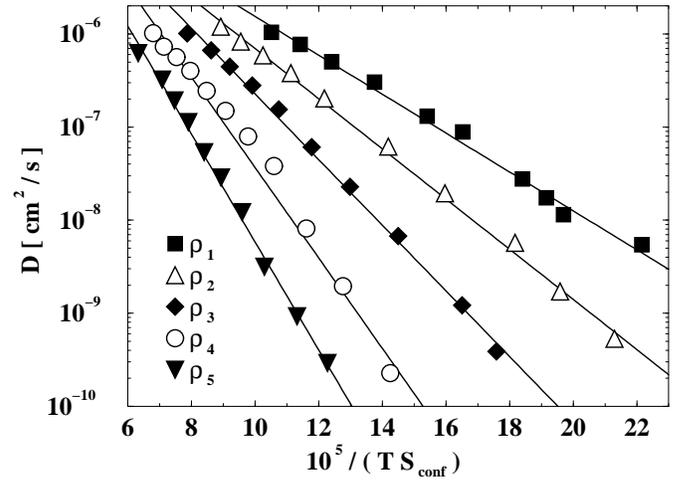}
\hfil}
\caption{Test of the Adam-Gibbs relation $\log D(T)\propto (1/T S_c)$ for five 
different densities. Note that this linear relation holds
both above and below the estimated critical temperatures $T_c$.}
\label{fig:adam_gibbs_fig}
\end{figure}

We finally study the link between configurational 
entropy and diffusion coefficient, investigating the 
validity of the Adam-Gibbs equation.
Fig.~\ref{fig:adam_gibbs_fig} shows $\log D$ as a
function of  $1/(T S_c)$; for all studied isochores,
$\log D$ vs. $1/(T S_c)$ is well described by a linear relation,
with coefficients which are volume dependent, as previously
found for the LJ liquid~\cite{sastry01},for the 
SPC/E model for water\cite{scala00} and for the BKS model for
silica\cite{voivod01}.

We note on passing that deviations from linear behavior are
observed at large value of $\log D$, where intra- and inter- basin
dynamics time scales are no longer separated. At high $T$,
it has been proposed\cite{zugutov} that 
entropy ---as opposed to configurational entropy--- is the
relevant thermodynamic quantity controlling dynamics.

\section{DISCUSSION AND CONCLUSIONS}
\label{conclusions}
In this article we have studied systematically the properties of the
potential energy surface for a simple three-site rigid model 
designed to mimic the properties of the fragile glass forming 
liquid ortho-terphenyl. The choice of this simple model, 
which collapses the entire phenyl ring into one
interaction site, allows us to run very long trajectories and to study 
in supercooled states the molecular
dynamics up to 1 $\mu s$, allowing the determination of diffusion
coefficients down to $10^{-10} cm^2/s$.

We have found that, as in the atomic LJ case, by cooling along an
isochore, basins of the PES of deeper and deeper energy are
explored. The basin volumes are functions of the depth in agreement
with previous studies.  Using the inherent structure
thermodynamic formalism, we have calculated the number of basins of
the PES and their depth, in the region of depth values probed by our
simulations.  As a result, we presented a full characterization
of the the temperature and density dependence of the basin depth,
degeneracy and volumes.

These results are used to provide a consistent model for the
intra-basin vibrational entropy.  This, together with the numerical
calculation of the total entropy via thermodynamic integration
starting from the ideal gas state, allow us to calculate the
configurational entropy --- the difference between the total entropy
and the vibrational one. This quantity is of primary interest both for
comparing with the recent theoretical calculations\cite{mezpar,coluzzi} 
and both to examine some of the proposed relation between dynamics and
thermodynamics\cite{adam65,wolynes,schultz} connecting a purely dynamical
quantity like the diffusion coefficient to a purely thermodynamical
quantity ($S_{c}$).  To examine such a possibility we compare
for five different isochores the $T$ dependence of $D$ with the
Adam-Gibbs relation.  In the entire range of $T$ and densities
studied the Adam Gibbs relation appears to provide a consistent
representation of the dynamics for the LW model. 

It is important to observe that a linear relation between 
$\log D$ and $1/(T S_c)$ holds both above and below
the ideal MCT critical temperature $T_c$, in agreement with similar
finding for the silica case\cite{voivod01}.
Recent works based on the instantaneous normal mode 
technique~\cite{keyes97} for several representative models
~\cite{stprl,donatiprl,lanaveprl,lanavelong,lanaveprlnew} 
provides evidence that above $T_{c}$ the system is always 
located in region of the PES close to the border between different basins. 
The number of diffusive directions 
significantly decreases above  $T_{c}$ and, if only
data above $T_{c}$ are considered, the 
number of  diffusive directions would appear to vanish at
$T_{c}$. Hence dynamics above $T_{c}$ is a dynamics of ``borders'' 
between basins and there is no clear reason 
why such dynamics should be well 
described by the Adam-Gibbs relation, which
focuses on the ``number'' of basins explored.  
The observed  validity of the AG relation ---both above and
below $T_{c}$--- reported in this manuscript 
supports the hypothesis that a direct relation exists between 
the number of basins and their connectivity\cite{lanaveprl,lanaveprlnew}. 
It is a challenge for future studies to confirm or disprove 
this hypothesis.
\begin{center}
{\bf ACKNOWLEDGEMENTS}
\end{center}
\noindent 
We thanks W.~Kob for very useful discussions.
We thanks INFM-PRA-HOP, INFM-Iniziativa Calcolo Parallelo, 
MIURST-COFIN-2000, and NSF Chemistry Program.
\begin{center}
{\bf APPENDIX A: IDEAL GAS ENTROPY FOR THE LW MODEL}
\end{center}
In this appendix we calculate the partition function of a
system of $N$ LW molecules in the non-interacting --ideal gas-- case.
The LW OTP molecule~\cite{lewis} is a rigid 
three sites isosceles triangle; each site represents
an entire phenyl ring of mass $m=6 m_C\simeq 78$ a.m.u.,
where $m_C$ is the mass of the carbon atom,   
and it is considered as the LJ interaction site. 
The length of the two short sides of the triangle 
is $\sigma=0.483$ $nm$ and the angle between them is
$\theta=5\pi / 12$ ($75$ degrees). 

The three moments of inertia for the single molecule are:
\begin{eqnarray}
I_x&=&\frac{2}{3} m \sigma^2 \cos^2 
\left (\frac{\theta}{2}\right )=  1.248\times 10^{-44} \; kg \; m^2,   \nonumber \\
I_y&=&2 m\sigma^2\sin^2\left (\frac{\theta}{2}\right )= 2.204\times
10^{-44} \; kg \;m^2,
\end{eqnarray}
and
\begin{eqnarray}
I_z&=&m\sigma^2\left [\frac{2}{3}\cos^2\left (\frac{\theta}{2}\right
)+2 \sin^2\left (\frac{\theta}{2}\right )\right ] =3.452\times 10^{-44} 
\; kg \; m^2. \nonumber
\end{eqnarray}
We define the following quantities
\begin{equation}
{\cal A} \equiv \frac{6\pi m k_B}{h^2} \;\;\;,\;\;\;
{\cal R}_\mu\equiv\frac{8\pi^2 k_B I_\mu}{h^2}
\end{equation}
where $\mu$ denotes x, y, or z.
The translational and rotational partition
functions for the single molecule are, respectively~\cite{fotocopie}
\begin{eqnarray}
{\mathcal Z}_T(T,V)&=& V \sqrt{({\cal A}T)^3}\\
{\mathcal Z}_R(T,V)&=& \frac{1}{2}\sqrt{\pi}\sqrt{{\cal R}_x {\cal R}_y {\cal R}_z T^3},
\end{eqnarray}
so the total partition function for an ideal gas of OTP molecules 
can be expressed as
\begin{equation}
{\mathcal Z}_{id}(T,V,N)=\frac{({\mathcal Z}_T {\mathcal Z}_R)^N}{N!}.
\end{equation}
We approximate $N!\approx N^N e^{-N}$.
The free energy $F_{id}$ and the entropy $S_{id}$
of the non-interacting system then become
\begin{eqnarray}
F_{id}(T,V,N)= - k_B T \ln \left [{\mathcal
Z}_{id}(T,V,N)\right ] =&&\nonumber \\
N\left [ \frac{1}{2}\ln 2 + \ln V \sqrt{{\cal A}^3
{\cal R}_x {\cal R}_y {\cal R}_z} + 3\ln T - \ln N +1 \right]&&\\
S_{id}(T,V,N)=-\frac{1}{k_B}\frac{\partial}{\partial T} F_{id}(T,V,N)&&\nonumber \\
N k_B \left\{ 4+\frac{1}{2}\ln \pi-ln 2+\ln\left [
\frac{V\sqrt{{\cal A}^3{\cal R}_x {\cal R}_y {\cal R}_z}}{N} T^3\right ]\right \}&&
\end{eqnarray}
where the term $\ln 2$ is due to the two possible degenerate
angular orientations of the molecule~\cite{fotocopie}.
%
%
%

%
%
%
\newpage
\begin{table}
\centering
\begin{tabular}{||c|c|c|c||}
$k$  & $\rho_k$  ( g / cm$^3$ )& $V_k$ ( nm$^3$ ) & $L_k$ ( nm ) \\\hline\hline
1 & 1.036 & 126.647 & 5.022 \\
2 & 1.060 & 123.883 & 4.985 \\
3 & 1.083 & 121.120 & 4.948 \\
4 & 1.108 & 118.356 & 4.910 \\
5 & 1.135 & 115.593 & 4.871
\end{tabular}
\vskip 0.5cm
\caption{Densities, volumes and simulation box lengths calculated.}
\label{numerics1:table}
\end{table}
\begin{table}
\centering
\begin{tabular}{||c|c|c|c|c||}
$\rho_1$  & $\rho_2$ & $\rho_3$ & $\rho_4$ & $\rho_5$ \\\hline\hline
170   & 190   & 230   & 280  & 320  \\
185   & 200   & 240   & 300  & 340  \\
190   & 210   & 260   & 320  & 360  \\
195   & 230   & 280   & 340  & 380  \\
210   & 250   & 300   & 360  & 400  \\
220   & 280   & 320   & 380  & 420  \\
240   & 300   & 340   & 400  & 440  \\
260   & 320   & 360   & 420  & 460  \\
280   & 340   & 380   & 440  & 480  \\
300   & 360   & 410   & 460  & 530  \\
-     & -     & -     & 480  & -    
\end{tabular}
\vskip 0.5cm
\caption{Temperatures (in K) for which calculations are performed.}
\label{numerics2:table}
\end{table}
\begin{table}
\centering
\begin{tabular}{||c|c|c||}
$i$ & $a_i$ (MPa nm$^{3(i+1)}$) &  $p_i$ (MPa nm$^{3(i+1)}$)\\\hline\hline
1 & 4835.96272 $\times 10^3$  &  15943.2    \\
2 & 1000.53765 $\times 10^6$  & -256.591    \\
3 & 9654.69470 $\times 10^6$  &  1.1745     \\
4 & 3873.87001 $\times 10^{10}$ & -0.00111551 
\end{tabular}
\vskip 0.5cm
\caption{Fitting coefficients for the excess pressure as a function of
$1/V$ at $T=5000~K$ and at $T=380$ K }
\label{fit_press:tab}
\end{table}
\begin{table}
\centering
\begin{tabular}{||c|c||}
$k$ & $S (T^*)$ $J / ( mol K )$ \\\hline\hline
1 & 192.80\\
2 & 188.21\\
3 & 183.54\\
4 & 177.95\\
5 & 172.12
\end{tabular}
\vskip 0.5cm
\caption{Total entropy at five densities for the reference temperature $T^*$.}
\label{total_ref:tab}
\end{table}
\begin{table}
\centering
\begin{tabular}{||c|c|c|c|c||}
$\rho_k$ & $U_0$ (kJ / mol)  & $\alpha$ (kJ T$^{-3/5}$/ mol)&
$A (kJ / mol)$ & $B$ (kJ T / mol)\\\hline\hline
1 & -86.30 &  0.4385 & -79.11 & -285 \\
2 & -88.94 &  0.4716 & -80.14 & -436 \\
3 & -92.07 &  0.5231 & -81.88 & -676 \\
4 & -95.23 &  0.5762 & -81.36 & -965 \\
5 & -96.06 &  0.5731 & -81.89 & -1100
\end{tabular}
\vskip 0.5cm
\caption{First two coloumns are the coefficients for the potential
energy $U(T,V)=U_0(V)+\alpha(V) T^{3/5}$; 
second two columns are the coefficients for the inherent structures
$e_{IS}(V,T)=A(V)+B(V)/T$.}
\label{energy_fit:tab}
\end{table}
\begin{table}
\centering
\begin{tabular}{||c|c|c||}
$\rho_k$ & $a (V) $  & $b(V) ( mol / kJ )$ \\\hline\hline
1 &47.1& 0.342\\
2 &41.2& 0.259\\
3 &36.5& 0.192\\
4 &32.1& 0.132\\
5 &28.9& 0.869
\end{tabular}
\vskip 0.5cm
\caption{Coefficients of the fit to the form 
$N^{-1}\sum_{k=1}^{6N-3} \log(\omega_k/\omega_o)=a(V)+b(V) \; e_{IS}(T)$.
}
\end{table}
%
%
\begin{figure}
\hbox to\hsize{\epsfxsize=1.0\hsize\hfil\epsfbox{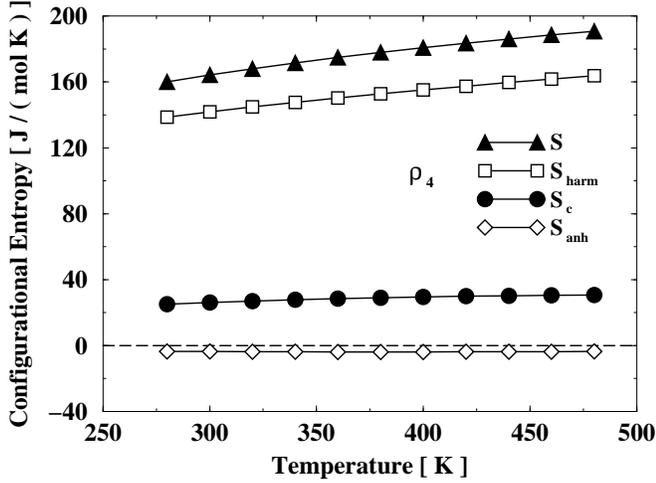}
\hfil}
\caption{Temperature dependence of the different contributions 
to the total entropy (closed triangles)
at the fixed selected density $\rho_4$: harmonic (open
squares), configurational entropy (closed circles) 
and anharmonic (open diamonds).}
\label{fig:s_2.90_fig}
\end{figure}
\end{multicols}
\end{document}